\documentclass [a5paper,10pt,article,oneside]{memoir}
\usepackage{pages}
\acinq
\aquatre

\usepackage[english]{babel}
\usepackage[utf8]{inputenc}
\usepackage{amsmath}
\usepackage{amsfonts,mathrsfs,paralist,amssymb,bm,amsthm}
\usepackage{graphicx,color,paralist}

\newlength{\tempa}

\newlength{\temp}
\newlength{\hauttrans}
\newlength{\longtrans}
\newlength{\longplace}

\newcommand{\petriexec}{%
\setlength{\hauttrans}{.66ex}
\setlength{\longtrans}{6ex}
\setlength{\longplace}{4ex}
\settoheight{\temp}{\strut}
\settodepth{\tempa}{\strut}
\addtolength{\temp}{\tempa}
\setlength{\temp}{2\temp}
\addtolength{\temp}{\hauttrans}}

\newcommand{\transh}{*+[F]\txt{\vbox to\hauttrans{\hbox to\longtrans{}}}}
\newcommand{\transhp}{*+[F--]\txt{\vbox to\hauttrans{\hbox to\longtrans{}}}}
\newcommand{\transv}{*+[F]\txt{\vbox to\longtrans{\hbox to\hauttrans{}}}}
\newcommand{\transvp}{*+[F--]\txt{\vbox to\longtrans{\hbox to\hauttrans{}}}}
\newcommand{\place}{*+[o][F]{\vbox to\longplace{\hbox to\longplace{}}}}
\newcommand{\transhlr}[1]{\POS[]\POS!R\drop{\strut\hspace{2ex}\rlap{$#1$}}}

\newcommand{\transhll}[1]{\POS[]\POS!L\drop{\llap{$#1$}\hspace{2ex}}}

\newcommand{\transhld}[1]{\POS[]\POS!D\drop{\raisebox{-1.5\temp}{\strut $#1$}}}

\newcommand{\transht}{*+[F.]\txt{\vbox to\hauttrans{\hbox to\longtrans{}}}}
\newcommand{\placet}{*+[o][F.]{\vbox to\longplace{\hbox to\longplace{}}}}

\newcommand{\markp}{\POS[]\drop{\bullet}}

\tiny\petriexec
\normalsize
\usepackage{hyperref}
\hypersetup{colorlinks=false,hidelinks=true}
\usepackage{mathrsfs}
\usepackage[all]{xy}
\xyoption{poly}
\newcommand{\addpoint}[1]{#1\ ---\ }

\setsecnumdepth{subsubsection}

\setsecnumformat{\csname the#1\endcsname---}
\setsubsechook{\setsecnumformat{\csname the##1\endcsname\ ---\ }}
\setsechook{\setsecnumformat{\csname the##1\endcsname---}}

\setsecheadstyle{\centering\Large\bfseries\sffamily}
\setsubsecheadstyle{\large\bfseries\sffamily}
\setsubsubsecheadstyle{\itshape\addpoint}
\setsubsubsecindent{1em}
\setbeforesubsubsecskip{0em}
\setsubsubsechook{\setsecnumformat{{\normalfont\csname
      the##1\endcsname\ }}}
\setaftersubsubsecskip{-0em}
\setsubparaheadstyle{\itshape}
\newtheoremstyle{thm}
     {1.5ex plus .3ex minus .1ex}
     {1ex plus .3ex minus .1ex}
     {\itshape}
     {}
     {\bfseries\sffamily}
     {---}
     {0em}
     {$\bullet$\hbox{\ }#1\hbox{\ }#2}

\theoremstyle{thm}

\newtheorem{definition}{Definition}[section]

\newtheorem{lemma}[definition]{Lemma}

\newtheoremstyle{note}
     {1ex plus .3ex minus .1ex}
     {1ex plus .3ex minus .1ex}
     {}
     {}
     {\itshape}
     {.}
     {1em}
     {}
\theoremstyle{note}

\newlength{\remaining}

\newcommand{\tq}{\;:\;}

\def\calli#1{\expandafter\def\csname
  #1\endcsname{\mathcal{#1}}}	
\def\sets#1{\expandafter\def\csname
  bb#1\endcsname{\mathbb{#1}}}	
\def\rebar#1{\expandafter\def\csname #1bar\endcsname{\overline{\csname
      #1\endcsname}}}		
\def\gothify#1{\expandafter\def\csname
  #1#1#1\endcsname{\mathfrak{#1}}}	

\sets{R}	
\sets{N}	
\calli{I}	
\calli{M}	
\calli{T}	
\calli{V}
\calli{U}
\calli{W}
\gothify{F}	
\gothify{G}	
\rebar{M}	

\newcommand{\C}{\mathscr{C}}

\newcommand{\up}[1]{\,\uparrow #1}

\newcommand{\Cstar}{\mathfrak{C}}
\newcommand{\slgb}{\mbox{$\sigma$-al}\-ge\-bra}

\newcommand{\BM}{\partial\M}

\newcommand{\R}{\mathscr R}

\newcommand{\unit}{\bm\varepsilon}

\newcommand{\pre}[1]{{}^{\bullet}#1}
\newcommand{\post}[1]{#1^\bullet}

\tightlists

\newlength{\firstl}
\newlength{\secondl}
\newlength{\finall}

\newcommand{\firstc}{Toward Uniform Random}
\newcommand{\secondc}{Generation in 1-safe Petri Nets}
\newcommand{\format}[1]{\makebox[\finall][s]{#1}} \newsavebox{\firstb}
\newsavebox{\secondb}

\newcommand{\lfaast}{%
  \savebox{\firstb}{\firstc}%
  \savebox{\secondb}{\secondc}%
  \setlength{\firstl}{\wd\firstb}%
  \setlength{\secondl}{\wd\secondb}%
  \ifnum\firstl>\secondl\setlength{\finall}{\firstl}\else
  \setlength{\finall}{\secondl}\fi\par
  \format{\firstc}\\
\format{\secondc}%
  }

\everymath{\normalfont}

\begin{document}
\mainmatter
\strut\vspace{-4em}
\begin{center}
\huge\bfseries\sffamily
\lfaast\par
\normalfont
\end{center}

\bigskip

\begin{center}
\begin{tabular}{c}
  \Large\sffamily Samy Abbes\\
  \small University Paris Diderot -- Paris~7\\
  \small CNRS Laboratory IRIF (UMR 8243)\\
  \small Paris, France\\
  \small
  \ttfamily\footnotesize  samy.abbes@univ-paris-diderot.fr
\end{tabular}
\end{center}

\begin{abstract}
  We study the notion of uniform measure on the space of infinite
  executions of a 1-safe Petri net. Here, executions of 1-safe Petri
  nets are understood up to commutation of concurrent transitions,
  which introduces a challenge compared to usual transition
  systems. We obtain that the random generation of infinite executions
  reduces to the simulation of a finite state Markov
  chain. Algorithmic issues are discussed.
\end{abstract}

\section{Introduction}
\label{sec:introduction}

Petri nets are formal models designed to describe and analyze the
behavior of concurrent systems. Among the many kinds of systems where
Petri nets may be introduced to formally describe a concurrent
dynamics, distributed databases \cite{diekert90} and telecommunication
networks \cite{benveniste03} are two typical examples. Both examples
involve temporal evolution on the one hand, and the paradigm of
resource sharing on the other hand, where resources are ``spatially''
distributed.  Requests for resources are local, in such a way that any
two actions requiring disjoint sets of resources may be considered as
parallel.

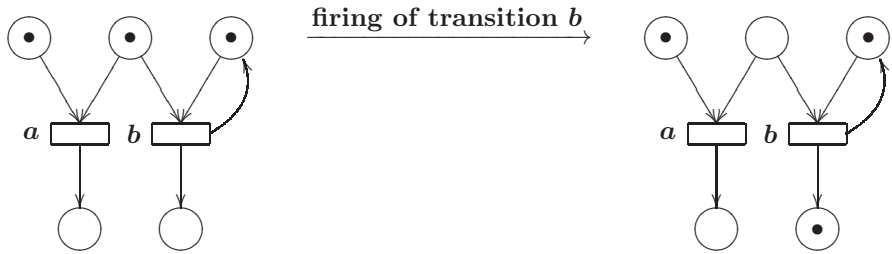
\begin{figure}
\begin{align*}
&
\xymatrix@C=0pt{  \place\ar[dr]!U\markp&&\place\ar[dl]!U\ar[dr]!U\markp&&\place\ar[dl]!U\markp\\
&\transh\ar[d]\transhll{\bm
  a}&&\transh\ar[d]\POS!R\ar@/_1em/[ur]\transhll{\bm b}\\
&\place&&\place
}
&
\xrightarrow{\text{\normalsize\bfseries firing of transition $\bm b$}}&&
\xymatrix@C=0pt{  \place\ar[dr]!U\markp&&\place\ar[dl]!U\ar[dr]!U&&\place\ar[dl]!U\markp\\
&\transh\ar[d]\transhll{\bm a}&&\transh\ar[d]\POS!R\ar@/_1em/[ur]\transhll{\bm b}\\
&\place&&\place\markp
}
\end{align*}
\caption{\textsl{Illustrating the firing rule in Petri nets}}
  \label{fig:firibngrule}
\end{figure}

Since their initial introduction in the 1960's, several variants of
Petri nets have been studied. In this paper, we shall limit ourselves
to 1-safe Petri nets, which we briefly define now. An \emph{unmarked
  Petri net} is a triple $N=(P,T,F)$, where $P$ and $T$ are two finite
and disjoint non empty sets of \emph{places} and of \emph{transitions}
respectively, and $F\subseteq (P\times T)\cup(T\times P)$ is called
the \emph{flow relation}. Graphically, places are traditionally
represented by circles and transitions are represented by squares or
rectangles (see Figure~\ref{fig:firibngrule}). The flow relation is
depicted by arrows from places to transitions and from transitions to
places.

Given a transition $t\in T$, the \emph{preset} $\pre t$ and the
\emph{postset} $\post t$ of~$t$ are the sets of places defined as
follows:
\begin{align*}
  \pre t&=\{p\in P\tq (p,t)\in F\}\,,&
                                       \post t&=\{p\in P\tq(t,p)\in F\}\,.  
\end{align*}
It is assumed that the preset and the postset of any transition are
both non empty.

A \emph{marking} of $N$ is any integer valued function $M:P\to\bbN$.
The marking is said to be \emph{$1$-safe}, or simply \emph{safe},
whenever $M(p)\leq1$ for all places $p\in P$. The number $M(p)$ is
interpreted as a number of \emph{tokens} lying in the
place~$p$. Tokens are graphically figured inside places, as in
Figure~\ref{fig:firibngrule}. Given a marking $M$ of~$N$, and a
transition $t\in T$, we say that $t$ can \emph{fire from~$M$}, or that
\emph{$M$~enables~$t$}, whenever $M(\cdot)>0$ on~$\pre t$\,.

If $t$ can fire from~$M$, then the \emph{firing rule} $M\xrightarrow t
M'$ defines the new marking $M'$ as follows (see the commentary below
and the illustration in Figure~\ref{fig:firibngrule}):
\begin{gather*}
  \forall p\in P\quad M'(p)=
  \begin{cases}
M(p),&\text{if $p\notin(\pre
  t\cup\post t)$}\\    
M(p)-1,&\text{if $p\in\pre t\setminus\post t$}
\\
M(p)+1,&\text{if $p\in\post t\setminus\pre t$}\\
M(p),&\text{if $p\in(\pre t\cap\post t)$}
  \end{cases}
\end{gather*}

The intuitive interpretation of the firing rule is as follows. The
tokens located in places of the preset $\pre t$ are \emph{resources},
needed for firing and consumed by the transition when fired; this
explains the second rule. The firing of $t$ also produces new
resources, \emph{i.e.}, tokens; the third rule specifies that the new
tokens are created in the postset of~$t$. If a place belongs to both
the pre- and the postset of~$t$, then we can see the fourth rule as
the simultaneous instance of both the second and the third rule:
an existing resource is consumed and immediately after a new one is
created, replacing the first one. Finally, the first rule specifies
that only places ``near'' the firing transition are concerned by the
firing rule. Note that the total number of tokens in the net may
differ once the transition has fired.

Given a marking~$M_0$\,, a \emph{firing sequence from~$M_0$} is a
sequence $(t_1,\ldots,t_k)$ of transitions such that, for some
markings $M_1,\ldots,M_k$ defined inductively, the firing rules
$M_{i-1}\xrightarrow{t_i}M_{i}$ hold for all $i=1,\ldots,k$\,. Of
course, if such markings exist, they are unique. Retaining only the
last marking, we introduce the obvious notation
$M_0\xrightarrow{t_1\cdots t_k}M_k$\,, with the convention
$M_0\xrightarrow{\varepsilon}M_0$ for the empty
sequence~$\varepsilon$, which is considered as a firing sequence.

Any marking $M$ such that $M_0\xrightarrow{t_1\cdots t_k}M$ holds for
some firing sequence $t_1\cdots t_k$ is said to be \emph{reachable
  from~$M_0$}\,. 

A \emph{Petri net} is a quadruple $N=(P,T,F,M_0)$, such that $(P,T,F)$
is an unmarked Petri net, of which $M_0$ is a marking, called the
\emph{initial marking of the net}. The Petri net $N$ is said to be
\emph{$1$-safe}, or simply \emph{safe}, whenever all markings
reachable from $M_0$ are safe; this implies of course that $M_0$
itself is safe. By convention, we assume that $M_0$ is fixed once and
for all, and by a \emph{marking} $M$ of the net we mean any marking
reachable from~$M_0$\,.

Based on firing sequences, authors have introduced a refined point of
view on the executions of Petri nets. The idea is to underline the
\emph{concurrency features} of the model. Indeed, the firing rule
enlightens that a special status should be given to pairs $(t,t')$ of
transitions such that
$(\pre t\cup\post t)\cap(\pre {t'}\cup\post{t'})=\emptyset$; in this
paper, we shall say that two such transitions are
\emph{distant}. According to the firing rule, if both $t$ and $t'$ are
enabled by some marking~$M$, and if they are distant,
then: \begin{inparaenum}[1)]\item the firing of one does not prevent
  the firing of the other one, and
\item the markings resulting from the two firing sequences $t\cdot t'$
  and~$t'\cdot t$, fired from~$M$, are the same.
\end{inparaenum}

Henceforth, instead of mere firing sequences, it is natural to
consider \emph{equivalence classes} of firing sequences, with respect
to some congruence~$\R$, such that \mbox{$(t\cdot t',t'\cdot t)\in\R$}
for any two distant transitions $t$ and~$t'$\,. We define thus $\R$ as
the smallest congruence on the set firing sequences that contains all
pairs $(t\cdot t',t'\cdot t)$, for $t$ and $t'$ two distant
transitions. Hence, two firing sequences are congruent whenever one
can pass from one to another by applying a finite number of times
elementary transformations of the form:
\begin{gather*}
  (x)\cdot t\cdot t'\cdot(y)\xrightarrow{} (x)\cdot t'\cdot t\cdot(y)
\end{gather*}
where $(x)$ and $(y)$ are any two words on the alphabet~$T$, and $t$
and $t'$ are two distant transitions.

Equivalence classes of firing sequences are called
\emph{configurations}; intuitively, they only capture the causal
relations between transitions, and they leave unspecified the
chronological relations between distant and unrelated transitions; see
for instance G.~Winskel's notion of event structure and of unfolding
of Petri net for a precise account on this
interpretation~\cite{nie81}.

In this paper, we are interested in the random sampling of
configurations of a given safe Petri net. We insist that it shall not
be confused with the random sampling of firing sequences. The later
ultimately reduces to the standard procedure of sampling in a large,
but finite transition system, with the set of markings of the net as
set of states. By contrast, the sampling of configurations is by no
means standard---at least, not on first sight.

Therefore, we consider configurations as our primary objects of
interest, instead of their underlying firing sequences. It is
consistent with the general idea that the concurrent firing of distant
transitions should not be ordered, since the ordering of parallel
transitions does not correspond to any ``observable'' of the system
and thus would be rather artificial.

In order to properly deal with configurations, we rely first on the
notion of \emph{trace monoid} (also called \emph{heap
  monoid}\/~\cite{viennot86} or \emph{free partially commutative
  monoid}\/~\cite{cartier69}). More precisely, the exact notion that
we need is that of a \emph{trace monoid acting on a finite
  set}. Associated with the action of a trace monoid, is a notion of
uniform probability measure on the space of ``infinite traces'',
developed in a previous work~\cite{abbes_dyn_2015}.  In the context of
Petri nets, infinite traces correspond to the infinite executions of
the net.

The purpose of this paper is thus twofold:
\begin{inparaenum}[1)]
\item give an account on the uniform measure relative to the action of
  a trace monoid on a finite set; and
  \item analyze the application of this theory in the context of safe
    Petri nets and for random sampling purposes.
\end{inparaenum} 
The intended applications related to random generation are of two
kinds: random generation of infinite executions of a Petri net,
uniformly distributed and once this has been properly defined, on the
one hand; random generation of a finite execution, uniformly
distributed among all finite executions of size~$n$, on the other
hand. The results of the paper allow to tackle the first task, at
least theoretically, although some algorithmic issues inevitably
remain and are discussed. The second task, concerning finite sampling,
is not discussed here because of space constraints. Let us mention
however that the solution to a similar, but simpler problem treated
in~\cite{abbes15:_unifor}, could probably be extended to this case
with few modifications. In turn, these are the basic building blocks
needed, for instance, for probabilistic model checking of formal
properties of concurrent systems.

The application to Petri nets will be treated on an
example. Henceforth, the contributions of this paper consist mainly in
the new questions that we formulate concerning the uniform measure
relatively to the specific model of Petri nets, compared to more
general asynchronous systems.

We underline that the probabilistic dynamics to which we are naturally
brought radically differs from that of \emph{stochastic Petri
  nets}~\cite{baccelli92:_synch_linear}---this is due to the radical
turn we have taken in our analysis by considering configurations
instead of firing sequences, as explained above. 

The outline of the paper is as
follows. Section~\ref{sec:trace-monoids} recalls the basics on trace
monoids and their combinatorics. Section~\ref{sec:acti-trac-mono} is
devoted to the action of trace monoids and their combinatorics, and
introduces the associated notion of uniform measure. It is only in
Section~\ref{sec:application-1-safe} that we meet with Petri nets
again, applying the results from Section~\ref{sec:acti-trac-mono} to
their case.

\section{Trace monoids and their boundary}
\label{sec:trace-monoids}

Let $\Sigma$ be an \emph{alphabet}, that is to say, a finite set, the
elements of which are called \emph{letters}. Let also
$I\subset\Sigma\times\Sigma$ be a binary relation, that we assume to
be symmetric and irreflexive, and that is called an \emph{independence
  relation}. Let $\Sigma^*$ denote the monoid of $\Sigma$-words, and
let $\R$ be the smallest congruence on $\Sigma^*$ containing all pairs
$(ab,ba)$ for $(a,b)\in I$.  The \emph{trace monoid} $\M=\M(\Sigma,I)$
is the quotient monoid $\M=\Sigma^*/\R$\,. In other words, $\M$~is the
presented monoid:
\begin{gather*}
  \M=\langle\Sigma\;|\;ab=ba\text{ for all $(a,b)\in I$}\rangle\,.
\end{gather*}

Elements of $\M$ are called \emph{traces}. The concatenation of traces
is denoted with the dot~``$\cdot$'', and the identity element is
denoted~$\unit$.  Clearly, if $x$ and $y$ are two $\R$-congruent
words, they have the same length. This defines a length function
$|\cdot|:\M\to\bbN$ on traces. Clearly also, the left divisibility
relation on~$\M$, defined by $x\leq y\iff\exists z\quad y=x\cdot z$,
is a partial order on~$\M$. 

A \emph{clique} of $\M$ is any product of the form
$\gamma=a_{1}\cdot\ldots\cdot a_{k}$\,, where $a_1,\ldots,a_k$ are
letters such that $i\neq j\implies(a_i,a_j)\in I$\,. Note in
particular that the $a_i$'s are pairwise distinct. Let $\C$ denote the
set of cliques; this is a finite set. Let also
$\Cstar=\C\setminus\{\unit\}$ denote the set of non empty cliques. A
pair $(\gamma,\gamma')\in\Cstar\times\Cstar$ is said to be in
\emph{normal form} whenever, for each letter $b$ occurring
in~$\gamma'$, there is a letter $a$ occurring in $\gamma$ such that
$(a,b)\notin I$. We denote it by the symbol $\gamma\to\gamma'$\,.

A sequence $(\gamma_1,\ldots,\gamma_n)$ in $\Cstar$ is said to be
\emph{normal} whenever $\gamma_i\to\gamma_{i+1}$ holds for all
$i\in\{1,\ldots,n-1\}$\,. It is well known that traces admit the
following normal form~\cite{cartier69,viennot86}: for every
$x\in\M\setminus\{\unit\}$, there exists a unique integer $n\geq1$ and
a unique normal sequence $(\gamma_1,\ldots,\gamma_n)$ in~$\Cstar$ such
that $x=\gamma_1\cdot\ldots\cdot\gamma_n$\,.

The existence of this normal form is the basis for establishing the
following combinatorial results. Let $\mu(z)$ be the polynomial,
called \emph{M\"obius polynomial}, and defined by:
\begin{gather*}
  \mu(z)=\sum_{\gamma\in\C}(-1)^{|\gamma|}z^{|\gamma|}\,.
\end{gather*}

Let also the growth series $G(z)$ be defined by:
\begin{align*}
  G(z)&=\sum_{k\geq0}\lambda(k)z^k\,,&
\lambda(k)&=\# S(k)\,,&
S(k)&=\{x\in\M\tq|x|=k\}\,.
\end{align*}

Then $G(z)$ is a rational series, inverse of the M\"obius polynomial:
$G(z)=1/\mu(z)$\,. Furthermore~\cite{goldwurm00,krob03,csikvari13},
$\mu(z)$~has a unique root of smallest modulus. This root,
say~$p_0$\,, is real and lies in $(0,1]$, and coincides with the
radius of convergence of the series $G(z)$. We shall extend in next
section this well known result in the framework of a trace monoid
acting on a finite set.

The existence of the normal form entails that traces are in bijection
with finite paths in the finite graph of non empty cliques
$(\Cstar,\to)$. It is thus natural to define \emph{infinite traces} as
infinite paths in the very same graph $(\Cstar,\to)$. Let $\BM$ denote
the set of infinite traces. The set $\BM$ is called the \emph{boundary}
of~$\M$. It is standard that the set $\BM$, equipped with
the natural topology, is metrisable and compact.

It is a bit less standard to extend the partial ordering relation
$\leq$ from $\M$ to~$\BM$. Let $\xi=(\gamma_1,\gamma_2,\ldots)$ be an
infinite trace, and let $x\in\M$. Then we put $x\leq\xi$ if and only
if $x\leq(\gamma_1\cdot\ldots\cdot\gamma_k)$ in $\M$ for all $k$ large
enough. And we define the \emph{visual cylinder}~$\up x$, as the
following subset of~$\BM$\,:
\begin{align*}
\up x&=\{\xi\in\BM\tq x\leq\xi\}\,.
\end{align*}

We will be interested in probability measures on the space $\BM$
equipped with its Borel \slgb. The following result shows the
importance of visual cylinders in this respect
(see~\cite[\S~2.2]{abbes15:_unifor_bernoul}).

\begin{lemma}
  \label{lem:1}
Any probability measure $m$ on\/ $\BM$ is
entirely characterized by the countable collection $m(\up x)$, for
$x$ ranging over\/~$\M$.
\end{lemma}

We conclude this section by illustrating the above notions on a
concrete example. It will be given a Petri net interpretation below in
Section~\ref{sec:application-1-safe}. 

Let $\Sigma=\{a,b,c,d,e\}$ and let
$\M=\langle\Sigma\;|\; ad=da,\ ae=ea,\ bd=db,\ be=eb,\
ce=ce\rangle$\,.
The set of non empty cliques is
$\Cstar=\{a,b,c,d,e,ad,ae,bd,be,ce\}$\,. For instance, the trace
$adebc=daebc=dabec$ has the
normal form $(ad)\to(be)\to(c)$.

The M\"obius polynomial of $\M$ is: $\mu(z)=1-5z+5z^2$\,, and
its root of smallest modulus is $p_0=\frac12-\frac1{2\sqrt5}$\,.

\section{Uniform measure relative to the action of a trace monoid}
\label{sec:acti-trac-mono}

Let $\M=\M(\Sigma,I)$ be a trace monoid.  Let also $Y$ be a finite set
of \emph{states}. We assume given a \emph{right monoid action} of $\M$
on~$Y$, that is to say, a function $\varphi:Y\times\M\to\M$ denoted by
$\varphi(s,x)=s\cdot x$, satisfying the following:
\begin{align*}
  \forall s\in Y\quad s\cdot\unit&=s\,,&
\forall (s,x,y)\in Y\times\M\times\M\quad s\cdot(x\cdot y)&=(s\cdot
                                                            x)\cdot y\,.
\end{align*}

The pair $(Y,\M)$, with the action understood, is called an
\emph{asynchronous system}.  The traces $x\in\M$ are thought of as
\emph{actions featuring parallelism}; the effect of the action
$x\in\M$ on a state $s\in Y$ is to change the state of the system into
the new state~$s\cdot x$\,.

A desirable feature is to disable some actions, depending on the
current state of the system. We render this feature by assuming the
existence of a particular state~$\bot$, such that $\bot\cdot
x=\bot$ for all $x\in\M$. We put $Y=X\cup\{\bot\}$ with
$\bot\notin X$, so that the ``real'' states are actually the
elements of~$X$; we will always restrict our attention to those pairs
$(s,x)\in X\times\M$ such that $s\cdot x\neq\bot$. Hence, it is
customary to introduce the following notation:
\begin{gather*}
  \forall s\in X\quad \M_s=\{x\in\M\tq s\cdot x\neq\bot\}\,.
\end{gather*}

We say that the action is \emph{irreducible} if, for every pair
$(s,t)\in X\times X$ of states, there exists an action $x\in\M$ such
that $x\neq\unit$ and $t=s\cdot x$. We shall always assume that the
action under consideration is irreducible.

On the combinatorics side, the analogous of the growth series is the
following matrix of formal series:
\begin{align*}
  G&=(G_{s,t})_{(s,t)\in X\times X}\,,&
G_{s,t}(z)&=\sum_{x\in\M\tq x\cdot s=t}z^{|x|}\,.
\end{align*}

Note that each line of~$G$, say indiced by $s\in S$, sums up to the
growth series $G_s$ of the subset~$\M_s$\,, defined by:
\begin{align}
\label{eq:2}
  G_s(z)&=\sum_{k\geq0}\lambda_s(k)\,,&\lambda_s(k)&=\#\{x\in\M_s\tq|x|=k\}\,.
\end{align}

Let us define the M\"obius matrix $M$ by:
\begin{align*}
  M&=(M_{s,t})_{(s,t)\in X\times X}\,,&
M_{s,t}(z)&=\sum_{\gamma\in\C\tq s\cdot\gamma=t}(-1)^{|\gamma|}z^{\gamma}\,,
\end{align*}
where $\C$ denotes the set of cliques of~$\M$\,. Then $M(z)$ is the
formal inverse of the growth matrix: $G(z)M(z)=I$, where $I$ is the
identity matrix of size $|X|\times|X|$
(see~\cite[Theorem~5.10]{abbes_dyn_2015}).

Given our hypothesis that the action is irreducible, all entries in
the growth matrix~$G$ have the same radius of convergence, say
$q_0\in(0,1]$. It is also the common radius of convergence of the
growth series $G_s(z)$, for $s\in X$, defined in~(\ref{eq:2}).  We
call $q_0$ the \emph{characteristic root} of the system $(Y,\M)$. 

Let the \emph{theta polynomial} be the polynomial with integer
coefficients defined by $\theta(z)=\det M(z)$. Then $q_0$ is the
smallest positive root of~$\theta(z)$
\cite[Theorem~5.11]{abbes_dyn_2015}.

\medskip The above elements of combinatorics allow us to introduce a
notion of uniform measure, as follows. For each state $s\in X$, and
for each real number $q\in(0,q_0)$, let $m_{s,q}$ be the discrete
probability measure on $\M$ defined by:
\begin{gather*}
  m_{q,s}=\frac1{G_s(q)}\sum_{x\in\M_s}q^{|x|}\delta_{\{x\}}\,,
\end{gather*}
where $\delta_{\{x\}}$ is the Dirac measure at~$x$. Then, according
to~\cite[Theorem~5.16]{abbes_dyn_2015}, the family $(m_{s,q})_{q<q_0}$
converges weakly, as $q\to q_0^-$\,, toward a probability measure
$\nu_s$ concentrated on~$\BM$, and which satisfies the following
property (recall that a probability measure on $\BM$ is entirely
determined by its values on visual cylinders):
\begin{gather}
\label{eq:3}
  \forall s\in X\quad\forall x\in\M_s\quad \nu_s(\up
  x)=q_0^{|x|}\Gamma(s,s\cdot x)\,,
\end{gather}
and $\nu_s(\up x)=0$ whenever $x\notin\M_s$\,, \emph{i.e.}, whenever
$s\cdot x=\bot$\,. In~(\ref{eq:3}),
$\Gamma(\cdot,\cdot):X\times X\to (0,\infty)$ is a function satisfying
the following \emph{cocycle identity}:
$\Gamma(s,u)=\Gamma(s,t)\Gamma(t,u)$ for all $s,t,u\in X$\,.

We call the family $\nu=(\nu_s)_{s\in X}$ the \emph{uniform measure}
of the asynchronous system $(Y,\M)$, although it is only each
individual $\nu_s$ which is actually a probability measure. Clearly,
the state $s$ indexing $\nu_s$ is to be seen as the \emph{initial
  state} of the system. The form~(\ref{eq:3}), together with the
cocycle identity, immediately yields the following property:
\begin{gather}
\label{eq:4}
  \forall (s,x,y)\in X\times\M\times\M\quad\nu_s\bigl(\up(x\cdot
  y)\bigr)=\nu_s(x)\cdot\nu_{s\cdot x}(\up y)\,.
\end{gather}

This chain rule~(\ref{eq:4}) justifies that the uniform measure $\nu$
is called a \emph{Markov measure}; but there are other families
satisfying~(\ref{eq:4}) than the uniform measure that we just
constructed.

Regarding a uniqueness result, we leave the following question open:
is there a unique pair $(q_0,\Gamma)$, where $q_0\in(0,1]$ and
$\Gamma(\cdot,\cdot):X\times X\to(0,\infty)$ satisfies the cocycle
identity, such that the formula
$\nu_s(\up x)=q_0^{|x|}\Gamma(x,s\cdot x)$ for $x\in\M_s$ and
$\nu_s(\up x)=0$ for $x\notin\M_s$\,, defines a probability measure on
$\BM$ for all $s\in X$?

\medskip From the sampling point of view, the following
\emph{realization result} is interesting. Let $s_0\in X$ be a fixed
initial state, and let $\xi\in\BM$ be an infinite trace distributed
according to~$\nu_{s_0}$\,. Let $\xi$ be given as the infinite path
$\xi=(C_1,C_2,\ldots)$ in $(\Cstar,\to)$ (see
Section~\ref{sec:trace-monoids}). Here, we see $(C_k)_{k\geq1}$ as a
sequence of random variables under the
probability~$\nu_{s_0}$\,. Finally, for $k\geq0$, let $S_k$ be the
state of the system reached after the action of the $k$ first cliques:
$S_k=s_0\cdot(C_1\cdot\ldots\cdot C_{k})$\,, with $S_0=s_0$ by
convention. Then, according to \cite[Theorem~4.5]{abbes_dyn_2015},
under~$\nu_{s_0}$\,, the sequence of pairs $(S_{k-1},C_k)_{k\geq1}$ is
a homogeneous Markov chain with values in $(X\times \Cstar)$, that we
call the Markov chain of \emph{states-and-cliques}. Furthermore:
\begin{inparaenum}[1)]
\item the transition matrix of the chain is independent of~$s_0$\,;
  and
\item both the initial measure and the transition matrix of the chain
  have algebraic expressions involving only $q_0$
  and~$\Gamma(\cdot,\cdot)$.
\end{inparaenum}
See~\cite{abbes_dyn_2015} for the explicit expressions.

Hence, both for theoretical and for sampling purposes, the problem
comes down to determining the characteristic root $q_0$ and the
cocycle $\Gamma(\cdot,\cdot)$. As for~$q_0$\,, its characterization as
a particular root of the theta polynomial yields the following
algorithmic issues:
\begin{inparaenum}[1)]
\item list all cliques of~$\M$, which is very well known to be hard
  (exponential in the size of $\Sigma$ in worst case~\cite{tomita06});
\item compute a determinant of size $|X|\times|X|$, the entries of
  which are polynomials of degree at most~$\alpha$, where $\alpha$ is
  the maximal size of a clique;
\item solve the equation $\theta(z)=0$ for the smallest positive root,
  where $\theta$ is of degree at most~$\alpha|X|$ and $\alpha$ is the
  maximal size of a clique.
\end{inparaenum}

Regarding the cocycle $\Gamma(\cdot,\cdot)$\,, we have not found a
systematic way of determining it. We will see in next section how to
determine it in practice on an example. Based on the uniqueness
conjecture stated above, we can expect that this technique actually
works in general, yielding an algorithmic way of obtaining the
cocycle.

\section{Application to 1-safe Petri nets}
\label{sec:application-1-safe}

In this section, we study 1-safe Petri nets as asynchronous systems as
defined in the previous section. Although this point of view is quite
natural, it does not seem to have been adopted by other authors yet,
not formally at least.

Let $N=(P,T,F,M_0)$ be a 1-safe Petri nets. Define the set of states
$X$ as the state of reachable markings of~$N$. We consider $\Sigma=T$
as alphabet, and as independence relation~$I$, we take
$I=\{(t,t')\in T\times T\tq (\pre t\cup\post
t)\cup(\pre{t'}\cup\post{t'})=\emptyset\}$,
hence the set of distant pairs as defined in
Section~\ref{sec:introduction}. Finally, let $\M=\M(\Sigma,I)$ be the
associated trace monoid (see Section~\ref{sec:trace-monoids}).

Putting $Y=X\cup\{\bot\}$, we define an action of $\M$ on $Y$ as
follows. Let $M\in X$, and let $x\in\M$, of which $(t_1,\ldots,t_k)$
is a representative sequence of transitions. Then, if
$(t_1,\ldots,t_k)$ is a firing sequence from~$M$, we put $M\cdot x=M'$
such that $M\xrightarrow{t_1\cdot\ldots\cdot t_k}M'$\,; and if
$(t_1,\ldots,t_k)$ is not a firing sequence from~$M$, we put $M\cdot
x=\bot$. Obviously, this definition does not depend on the
representative sequence $(t_1,\ldots,t_k)$. It is clear also that
$M\cdot(x\cdot y)=(M\cdot x)\cdot y$ holds for all $x,y\in\M$. 

\begin{figure}
  \centering
\begin{gather*}
  \xymatrix@C=8pt{
&\place\ar[dl]!U\ar[dr]!U\markp\\
\transh\POS!U\ar@/^1em/[ur]\transhld{\bm
  a}&&\transh\ar[d]\transhlr{\bm b}\\
&&\place\ar[dr]!U&&\place\ar[dl]!U\ar[dr]!U\markp
&&\place\ar[dl]!U\ar[dr]!U\markp
\\
&&&\transh\POS!L\ar@/^2em/[uuull]!D\POS[]\POS!U\ar@/^1em/[ur]\transhld{\bm
  c}&&\transh\POS!L\ar@/^1em/[ul]\POS!R\ar@/_.25em/[ur]!D\transhld{\bm
  d}&&\transh\POS!L\ar@/^.5em/[ul]\transhld{\bm e}
}
\end{gather*}
\caption{\textsl{Example of a safe Petri net. The underlying trace
    monoid is the one introduced at the end of
    Section~\ref{sec:trace-monoids}.}}
  \label{fig:petrtopia}
\end{figure}
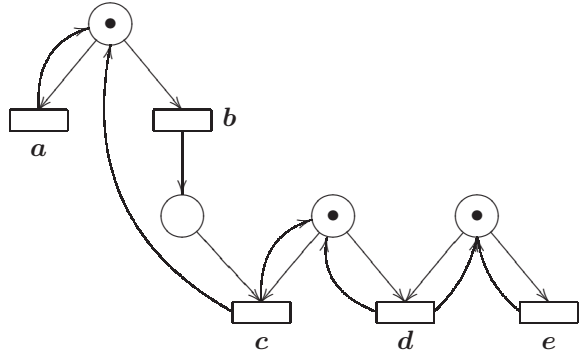

Let us consider for instance the Petri net depicted in
Figure~\ref{fig:petrtopia}, with the initial marking, say~$M_0$\,,
depicted. Clearly, this net is safe, and its only reachable marking,
say~$M_1$\,, other than $M_0$ is the marking obtained from $M_0$ by
firing transition~$b$. The associated trace monoid is
$\M=\langle a,b,c,d,e,\;|\; ad=da,\ ae=ea,\ bd=db,\ be=eb,\
ce=ec\rangle$,
that is to say, the example trace monoid introduced at the end of
Section~\ref{sec:trace-monoids}. The action obeys the following
description, which implies in particular that it is an irreducible
action:
\begin{align*}
  M_0\cdot a&=M_0&M_0\cdot b&=M_1&M_0\cdot c&=\bot&M_0\cdot
                                                    d&=M_0&M_0\cdot
                                                            e&=M_0\\
  M_1\cdot a&=\bot&M_1\cdot b&=\bot&M_1\cdot c&=M_0&M_1\cdot
                                                     d&=M_1&M_1\cdot e&=M_1
\end{align*}

Note however that this description does not characterize the action by
itself, since it does not render the concurrency features encoded in
the independence relation~$I$.

The M\"obius matrix and the theta polynomials are:
\begin{align*}
  M(z)&=
        \begin{pmatrix}
1-3z+2z^2&-z+2z^2\\
-z+z^2&1-2z
        \end{pmatrix}&
\theta(z)&=(1-z)(1-2z)(1-2z-z^2)\,.
\end{align*}
The characteristic root is thus $q_0=\sqrt2-1$. 

Let $\nu=(\nu_{M_0},\nu_{M_1})$ be the uniform measure associated with
the action. Let also $\Gamma(\cdot,\cdot)$ be the cocycle associated
with~$\nu$, so that $\nu_M(\up x)=q_0^{|x|}\Gamma(M,M\cdot x)$ for all
$M\in\{M_0,M_1\}$ and for all $x\in\M_M$\,. Then, by the cocycle
property of~$\Gamma$, one has $\Gamma(M_0,M_0)=\Gamma(M_1,M_1)=1$ on
the one hand, and one has $\Gamma(M_1,M_0)=1/\Gamma(M_0,M_1)$ on the
other hand. Hence $\Gamma(\cdot,\cdot)$ is entirely determined by the
single value $\lambda=\Gamma(M_0,M_1)$.

In order to retrieve the value~$\lambda$, we write down the following
equation. This is
equivalent~\cite{abbes15:_unifor_bernoul,abbes_dyn_2015} to writing
down the total probability law for the first clique~$C_1$\,, starting
from marking~$M_0$\,:
\begin{gather*}
  \sum_{\gamma\in\C\tq M_0\cdot\gamma\neq\bot}(-1)^{\gamma}\nu_{M_0}(\up\gamma)=0\,.
\end{gather*}

Since $\nu_{M_0}(\up \gamma)=q_0^{|\gamma|}\Gamma(M_0,M_0\cdot\gamma)$
on the one hand, and since the cliques $\gamma$ such that
$M\cdot\gamma\neq\bot$ range over $\{\unit,a,b,d,e,ad,ae,bd,be\}$,
this writes as:
\begin{gather*}
  1-3q_0\Gamma(M_0,M_0)-q_0\Gamma(M_0,M_1)+2q_0^2\Gamma(M_0,M_0)+2q_0^2\Gamma(M_0,M_1)=0\,.
\end{gather*}
Since $\Gamma(M_0,M_0)=1$, introducing the unknown
$\lambda=\Gamma(M_0,M_1)$ yields:
\begin{gather*}
\lambda=\frac{1-3q_0+2q_0^2}{q_0(1-2q_0)}=\frac{-1+3q_0}{1-2q_0}\,,
\end{gather*}
the later equality following from the equation $1-2q_0-q_0^2=0$.
Introducing the value $q_0=\sqrt2-1$ finally yields the simple
expression: $\lambda=\sqrt2$. Therefore the visual cylinders are given
probabilities with the following simple expressions:
\begin{align*}
\forall x\in\M_{M_0}\quad  \nu_{M_0}(\up x)&=
                                             \begin{cases}
                                               (\sqrt2-1)^{|x|},&\text{if
                                                 $M_0\cdot x=M_0$}\\
\sqrt2(\sqrt2-1)^{|x|},&\text{if $M_0\cdot x=M_1$}
                                             \end{cases}
\\
\forall x\in\M_{M_1}\quad\nu_{M_1}(\up x)&=
                                           \begin{cases}
(\sqrt2-1)^{|x|},&\text{if $M_1\cdot x=M_1$}\\
\frac1{\sqrt2}(\sqrt2-1)^{|x|},&\text{if $M_1\cdot x=M_0$}
                                           \end{cases}
\end{align*}

The above formulas however are not operational for sampling
purposes. One way to obtain a sampling method is to consider the
Markov chain of states-and-cliques, that is to say, the random
sequence of pairs $(M_{k-1},C_k)_{k\geq1}$\,, where $(C_k)_{k\geq1}$
is the sequence of cliques forming an infinite trace $\xi$ and $M_k$
is the marking reached after the~$k^\text{th}$~clique (see
Section~\ref{sec:acti-trac-mono}). Based on the formulas of
\cite{abbes_dyn_2015} and on the values found above for $q_0$ and
for~$\Gamma(\cdot,\cdot)$, we find that the transition matrix of
$(M_{k-1},C_k)_{k\geq1}$ is the following stochastic matrix---keeping
in mind that only the reachable pairs $(M,\gamma)\in X\times\Cstar$
have to be considered, see the remark below:
\begin{gather*}
\footnotesize
\hspace{-2cm}  \begin{array}{l}
M_0,a\\
M_0,c\\
M_0,ad\\
M_0,ae\\
M_0,ce\\
M_1,b\\
M_1,d\\
M_1,e\\
M_1,bd\\
M_1,be    
  \end{array}
  \begin{pmatrix}
-1+\sqrt2&0&0&0&0&2-\sqrt2&0&0&0&0\\
-2+\frac32\sqrt2&0&1-\frac12\sqrt2&0&0&3-2\sqrt2&0&0&-1+\sqrt2&0\\
-7+5\sqrt2&0&3-2\sqrt2&3-2\sqrt2&0&10-7\sqrt2&0&0&-4+3\sqrt2&-4+3\sqrt2\\
-7+5\sqrt2&0&3-2\sqrt2&3-2\sqrt2&0&10-7\sqrt2&0&0&-4+3\sqrt2&-4+3\sqrt2\\
-7+5\sqrt2&0&3-2\sqrt2&3-2\sqrt2&0&10-7\sqrt2&0&0&-4+3\sqrt2&-4+3\sqrt2\\
0&3-2\sqrt2&0&0&-2+\frac32\sqrt2&0&-1+\sqrt2&1-\frac12\sqrt2&0&0\\
0&3-2\sqrt2&0&0&-2+\frac32\sqrt2&0&-1+\sqrt2&1-\frac12\sqrt2&0&0\\
0&0&0&0&0&0&2-\sqrt2&-1+\sqrt2&0&0\\
0&3-2\sqrt2&0&0&-2+\frac32\sqrt2&0&-1+\sqrt2&1-\frac12\sqrt2&0&0\\
0&3-2\sqrt2&0&0&-2+\frac32\sqrt2&0&-1+\sqrt2&1-\frac12\sqrt2&0&0
\end{pmatrix}
\end{gather*}


Furthermore, the initial law of the chain, \emph{i.e.}, the law of
$(M_0,C_1)$\,, has the form $\delta_{\{M_0\}}\otimes\kappa$\,, where
$\kappa$ is the probability distribution on $\Cstar$ given by:
\begin{align*}
  \kappa(a)&=-7+5\sqrt2
&\kappa(b)&=10-7\sqrt2
&\kappa(c)&=0
&\kappa(d)&=0
&\kappa(e)&=0\\
\kappa(ad)&=3-2\sqrt2
&\kappa(ae)&=3-2\sqrt2
&\kappa(bd)&=-4+3\sqrt2
&\kappa(be)&=-4+3\sqrt2
&\kappa(ce)&=0
\end{align*}

A remark on these values: the equality $\kappa(c)=0$ is normal, since
$c$ is not enabled at~$M_0$\,. On the contrary, both $d$ and $e$ are
enabled at~$M_0$\,, and yet we find that $\kappa$ vanishes on $d$
and~$e$. This is obtained from a straightforward computation based on
the formulas of \cite{abbes_dyn_2015} applied to our case. But it is
also worth mentioning an intuitive explanation, as follows. Assume for
instance that $C_1=d$\,. Then the system stays in~$M_0$\,, and since
$c$ is not enabled at~$M_0$, the next clique will be either $d$
or~$e$, but will not contain either $a$ nor~$b$, by the definition of
the normal form. The same will happen for the next clique, and so
on. Henceforth $a$ and $b$ will never occur; but this has zero
probability, and it explains that $\kappa(d)=0$. The same holds for
$\kappa(e)=0$. The same reasoning also explains that the pairs
$(M_0,d)$ and $(M_0,e)$ are not reachable states of the Markov chain
of states-and-cliques, and thus do not appear as lines in the
transition matrix above.

Given the initial distribution and the transition matrix, one is of
course able to simulate the Markov chain
$(M_{k-1},C_k)_{k\geq1}$\,. In turn, the random trace
$(C_1\cdot\ldots\cdot C_k)$ corresponds to the $k$ first cliques of a
uniformly distributed infinite execution of the net.

We observe that the transition matrix displayed above has the
following property: for any clique~$\gamma$, the law of the next
marking starting from $(M_0,\gamma)$ is independent of~$\gamma$; and
similarly when starting from $(M_1,\gamma)$. We deduce that the
sequence $(M_k)_{k\geq0}$ is itself a Markov chain, with transition
matrix:
\begin{gather*}
  \begin{pmatrix}
    -1+\sqrt2&2-\sqrt2\\
1-\frac12\sqrt2&\frac12\sqrt2
  \end{pmatrix}
\end{gather*}

\medskip
Based on the results obtained for the above specific example, the
following questions arise naturally regarding how to deal with a
general 1-safe Petri net.
\begin{enumerate}
\item What would be an algorithmic way of computing the cocycle
  $\Gamma(\cdot,\cdot)$?
\item The uniform measure has the simple characterization
  \mbox{$\nu_s(\up x)=q_0^{|x|}\Gamma(s,s\cdot x)$}\,. By contrast,
  its realization as a Markov chain involves a set of states which
  grows, in worst case, exponentially fast with the number of places
  and with the number of transitions of the net. It is therefore
  natural to seek for other means for uniform generation than the
  computation of the Markov chain of states-and-cliques. This relates
  with the next point.
\item In our specific example, the uniform measure clearly shows a
  probabilistic independence between the two \emph{parallel choices}
  between $a$ over $b$ on the one hand, and between $d$ over $e$ on
  the other hand. Until which extend can we generalize this spatial
  independence property for a safe Petri net?  Does it take a more
  specific formulation for certain particular classes of safe Petri
  nets, such as free-choice nets for instance~\cite{best87}? Can we
  use this independence property to devise an efficient sampling
  algorithm for executions of nets?
\item For our specific example, we have observed that the sequence of
  markings $(M_k)_{k\geq0}$ reached by the Markov chain of
  states-and-cliques $(M_{k-1},C_k)_{k\geq1}$ is itself a Markov
  chain; this is not the case for general asynchronous systems. Can we
  obtain sufficient conditions on the structure of a 1-safe Petri nets
  for this to hold?
\end{enumerate}

\bibliographystyle{plain}
\bibliography{gascom.bib}


\end{document}